\documentclass[superscriptaddress,prl,twocolumn,nofootinbib]{revtex4}
\usepackage{amsmath,amssymb}
\usepackage{graphicx,epsfig}
\usepackage{slashed}
\usepackage{float}

\def\beq{\begin{equation}}
\def\eeq#1{\label{#1}\end{equation}}
\def\eeqn{\end{equation}}
\def\beqa{\begin{eqnarray}}
\def\eeqa#1{\label{#1}\end{eqnarray}}
\def\eeqan{\end{eqnarray}}
\def\CR{\nonumber \\ }
\def\leqn#1{(\ref{#1})}
\newcommand{\centeron}[2]{{\setbox0=\hbox{#1}\setbox1=\hbox{#2}\ifdim
\wd1>\wd0\kern.5\wd1\kern-.5\wd0\fi \copy0
\kern-.5\wd0\kern-.5\wd1\copy1\ifdim\wd0>\wd1
                                   \kern.5\wd0\kern-.5\wd1\fi}}
\newcommand{\ltap}{\>\centeron{\raise.35ex\hbox{$<$}}
                           {\lower.65ex\hbox{$\sim$}}\>}
\newcommand{\gtap}{\>\centeron{\raise.35ex\hbox{$>$}}
                           {\lower.65ex\hbox{$\sim$}}\>}
\newcommand{\gsim}{\mathrel{\gtap}}
\newcommand{\lsim}{\mathrel{\ltap}}

\begin{document}

\title{750 GeV Di-photon Excess and Strongly First-Order Electroweak Phase Transition}

\author{Maxim Perelstein}
\email{mp325@cornell.edu}
\author{Yu-Dai Tsai}
\email{yt444@cornell.edu}
\affiliation{Laboratory for Elementary Particle Physics, Cornell University,
Ithaca, NY 14850, USA}

\date{\today}

\begin{abstract}
A new scalar particle, coupled to photons and gluons via loops of vector-like quarks, provides a simple theoretical interpretation of the 750 GeV diphoton excess reported by the experiments at the Large Hadron Collider (LHC). In this paper, we show that this model contains a large, phenomenologically viable parameter space region in which the electroweak phase transition (EWPT) is strongly first-order, opening the possibility that electroweak baryogenesis mechanism can be realized in this context. A large coupling between the Higgs doublet and the heavy scalar, required for a strongly first-order EWPT, can arise naturally in composite Higgs models. The scenario makes robust predictions that will be tested in near-future experiments. The cross section of resonant di-Higgs production at the 13 TeV LHC is predicted to be at least 20 fb, while the Higgs cubic self-coupling is enhanced by 40\% or more with respect to its Standard Model (SM) value.
\end{abstract}

%\pacs{14.60.Pq, 98.80.Cq, 98.70.Vc}

\maketitle
{\em Introduction ---} Experiments at the Large Hadron Collider (LHC) have recently reported an intriguing hint of a new resonance in the diphoton channel with an invariant mass of approximately 750 GeV~\cite{ATLAS,CMS}. If true, such a resonance could not be explained within the Standard Model (SM) of particle physics. Many theoretical interpretations of this signal in terms of physics beyond the SM have already been proposed. A simple and compelling interpretation is to postulate the existence of a new scalar (spin-0) particle, $X$, with $m_X\approx 750$ GeV, which is singly produced as an $s$-channel resonance in the collision of gluons, and decays directly into a photon pair. This scenario was explored, for example, in Refs.~\cite{Knapen:2015dap,Buttazzo:2015txu,Franceschini:2015kwy,Ellis:2015oso,Falkowski:2015swt,Berthier:2015vbb,Craig:2015lra,Dolan:2016eki,Staub:2016dxq}. (Production via photon collisions is also possible, see {\it e.g.}~\cite{Fichet:2015vvy,Csaki:2015vek,Csaki:2016raa}, but will not be our focus in this work.) While $X$ itself does not have any SM quantum numbers, its coupling to photons and gluons can be mediated by loops of vector-like quarks (VLQ's). This simple structure can accommodate the observed diphoton rate without contradicting the constraints on new physics from the LHC Run 1 and elsewhere, and it arises naturally in many new physics scenarios, including little Higgs and composite Higgs models (see, for example, Refs.~\cite{Harigaya:2015ezk,Franceschini:2015kwy,Belyaev:2015hgo,Craig:2015lra}). 

The origin of the asymmetry between matter and antimatter in the universe is one of the major open questions in fundamental physics. Several mechanisms that could generate this asymmetry have been proposed. One of the most compelling scenarios, the electroweak baryogenesis (for review, see {\it e.g.} Ref.~\cite{Morrissey:2012db}), requires that the electroweak phase transition be strongly first-order. In the SM, the transition from electroweak-symmetric to electroweak-broken vacuum is known to be a crossover~\cite{Csikor:1998eu}, so that the electroweak baryogenesis scenario cannot be realized. However, it is well known that physics beyond the SM can change the nature of the transition. In particular, the presence of extra scalar particles, with significant interactions with the Higgs, can trigger a strongly first-order EWPT. In this paper, we investigate whether the new scalar $X$, postulated to explain the 750 GeV diphoton excess, can play this role. (For previous work on this subject, see Refs.~\cite{Ghoshal:2016jyj,Chao:2016aer}.)

{\em Model ---} We introduce a real scalar field $S$, which has no SM quantum numbers. To avoid large mixing between $S$ and the SM Higgs $H$, we impose a reflection $Z_2$ symmetry, under which $S\to -S$ and all SM fields, including $H$, are even. The tree-level scalar potential has the form
\beq
V_{\rm tree} = -\mu^2 |H|^2 + \lambda |H|^4 + \frac{\mu_S^2}{2}S^2+\frac{\lambda_S}{4}S^4+\lambda_{SH} S^2 |H|^2.
\eeq{sca_pot}
This is the most general renormalizable potential allowed by the SM gauge symmetries and the $Z_2$. Note that no other renormalizable interactions of $S$ with the SM fields are allowed. The tree-level values of the parameters known from experiment are $\mu \approx 88$ GeV, $\lambda\approx 0.13$, and 
\beq
\sqrt{\mu_S^2+\lambda_{SH} v^2}=750~{\rm GeV}, 
\eeq{Xmass}
where $v\approx 246$ GeV is the SM Higgs vacuum expectation value (vev). In our analysis, we will treat $\lambda_{SH}$ as a free parameter, and use~\leqn{Xmass} to fix $\mu_S$. The value of $\lambda_S$ has only a marginal effect on the quantities of interest to us, and we fix it at 1.0 throughout the analysis. 

In addition, we introduce a VLQ field, a Dirac fermion $Q$, with a Yukawa coupling to the $S$:
\beq
{\cal L} \subset \ m_Q Q_L^\dagger Q_R + y_Q S Q_L^\dagger Q_R +~{\rm h.c.}
\eeq{L_VLQ}   
We postulate that the left- and right-handed components of $Q$ have opposite charges under the $Z_2$ symmetry, for example $Q_L\to Q_L$ and $Q_R\to -Q_R$. Then, the Yukawa coupling is $Z_2$-invariant, and the symmetry is only broken softly by the VLQ mass term. We assume this breaking to be spontaneous, {\it i.e.} the VLQ mass term arises from a $Z_2$-breaking vev of an additional scalar field; in this case, the $Z_2$ is fully restored above the scale $m_Q$. We further assume that such spontaneous breaking does not generate tree-level $Z_2$-breaking terms in the scalar potential. Extended symmetry structure and field content are generally required to avoid such terms; see Appendix for an explicit example.

To mediate $S$ interactions with gluons and photons, $Q$ needs to be colored and electrically charged; for concreteness, we assume that SM quantum numbers of $Q$ are $({\bf 3}, {\bf 1})_{5/3}$. Such charge-$5/3$ quarks are common in composite Higgs models. In the parameter range of interest, the interactions of $S$ with gluons and photons can be expressed in the effective operator formalism, by integrating out the VLQ:
\beq
{\cal L}=\frac{e^2}{4v} c_\gamma SF_{\mu\nu}F^{\mu\nu}+\frac{g_s^2}{4v} c_g SG_{\mu\nu}G^{\mu\nu},
\eeq{L_int}  
where $F$ and $G$ are the $U(1)_{\rm em}$ and $SU(3)_c$ field strength tensors, and
\beq
c_\gamma = \frac{y_Q Q_Q^2 v}{2\pi^2m_Q},~c_g = \frac{y_Q v}{12\pi^2m_Q}. 
\eeq{cees}
The VLQ loops will also induce the couplings of $S$ to $Z\gamma$ and $ZZ$ pairs. 

Radiative corrections involving VLQ loops induce $Z_2$-violating terms in the scalar potential:
\beq
V_{\slash{\hskip-2mm Z_2}} = a_1 S + a_3 S^3 + \delta_1 S |H|^2. 
\eeq{noZ2pot}
The coefficients $a_1$ and $a_3$ are induced at the one-loop level. Using $m_Q$ as the ultraviolet cutoff in the loop integrals (since the $Z_2$ symmetry is restored above this scale) yields an estimate
\beq
a_1 = \frac{A y_Q m_Q^3}{8\pi^2},~a_3 = \frac{B y_Q^3 m_Q}{8\pi^2},
\eeq{a_estimates} 
where $A$ and $B$ are order-one coefficients that depend on the details of the sector responsible for spontaneous $Z_2$ violation at the scale $m_Q$. The value of $B$ has very small effect on our analysis, and for concreteness we set it to 1. We treat $A$ as a parameter; however, note that values of $A$ far below 1 require fine-tuning.   
The coefficient $\delta_1$ is first generated at the two-loop level, and is numerically negligible. 

The tadpole term in the potential~\leqn{noZ2pot} inevitably causes $S$ to get a vev, which in turn induces a mixing of $S$ and $H$. Working to linear order in the small $Z_2$-breaking parameters, we obtain
\beq
\langle S \rangle \approx - \frac{a_1}{m_X^2},~~~\theta_{SH} \approx\frac{2\lambda_{SH} \,v \, \langle S \rangle}{m_X^2-m_\phi^2}.
\eeq{s_vev}
Here $\theta_{SH}$ is the angle of rotation between the gauge eigenbasis $(S, h)$ and the mass eigenbasis $(\phi, X)$, with $m_\phi=125$ GeV and $m_X=750$ GeV:
\beqa
\phi &=& \cos \theta_{SH}\, h - \sin\theta_{SH}\, S, \CR
X &=&  \sin\theta_{SH}\, h + \cos \theta_{SH}\, S.
\eeqa{rotation}
The linearized analysis is applicable as long as $\langle S \rangle \ll v$, $\theta_{SH}\ll 1$. The parameter space region of interest for our analysis has $|\langle S \rangle|\lsim 20$ GeV, $|\theta_{SH}|\lsim 4\times 10^{-3}$, which justifies treating the $Z_2$ breaking as a small perturbation. 

With the assumptions outlined above, the model has three input parameters, $(\lambda_{SH}, m_Q, y_Q)$, in addition to the constant $A$. An alternative basis is $(\lambda_{SH}, \kappa, c_g)$, where $\kappa\equiv a_1^{1/3}$. In this basis, $A$ only affects the VLQ mass at each parameter point, but is otherwise irrelevant, making it a convenient choice. 

{\it LHC Phenomenology---} A combined fit to the event rates of the 750 GeV diphoton excess seen at ATLAS and CMS yields~\cite{Buttazzo:2015txu}
\beq
\sigma(pp\to X) \, {\rm Br}(X\to \gamma \gamma) = (4.4 \pm 1.1)~{\rm fb}.
\eeq{xsec} 
For illustration, we will use the central value in our numerical analysis. We use Eq.~\leqn{xsec} to fix $c_g$ in terms of the other two parameters, $\lambda_{SH}$ and $\kappa$. In $\sigma(pp\to X)$, we take into account QCD radiative corrections, using the NNLO K-factor following Ref.~\cite{Buttazzo:2015txu}. The observed diphoton spectrum has a slight preference for a non-trivial $X$ width of about 50 GeV; however, the statistical significance of this hint is small, and we do not use this information in this study. 

In addition to photon pairs, in our model $X$ can decay to $gg$, $ZZ$ and $\gamma Z$ pairs through VLQ loops. The singlet-doublet mixing also induces Higgs-like decay modes such as $X\to t\bar{t}, W^+W^-$, and provides an additional (in fact, dominant) contribution to $X\to ZZ$. Finally, the $S$ vev as well as the mixing contribute to the decay $X\to \phi\phi$. Rates in all these channels are constrained by searches at the LHC Run-I; the strongest constraints on our model are imposed by the ATLAS searches in the di-Higgs~\cite{Khachatryan:2015yea,Aad:2015xja}, $W^+W^-$~\cite{Khachatryan:2015cwa,Aad:2015agg}, and $ZZ$~\cite{Khachatryan:2015cwa,Aad:2015kna} channels. The bounds on these three channels are 35 fb, 38 fb, and 22 fb, respectively. In addition, the production and decay rates of the 125 GeV state, $\phi$, are affected by the doublet-singlet mixing. The fractional deviations of the couplings of the 125 GeV state to photons and gluons from their SM values are given by
\beq
\kappa_\gamma \approx 1 + 12.0 \, c_\gamma \theta_{SH},~~~\kappa_g \approx 1 - 118 \, c_g \theta_{SH}.
\eeq{h_shifts}
Shifts in all other couplings are much smaller, being either quadratic in the mixing angle $\theta_{SH}$, or additionally suppressed by loop factor(s). ATLAS and CMS fits to the Higgs data in the ``loop-only new physics" framework~\cite{Khachatryan:2014jba,Aad:2015gba} can then be used to place constraints on our scenario.

\begin{figure}[t!]
\begin{center}
\includegraphics[width=8cm]{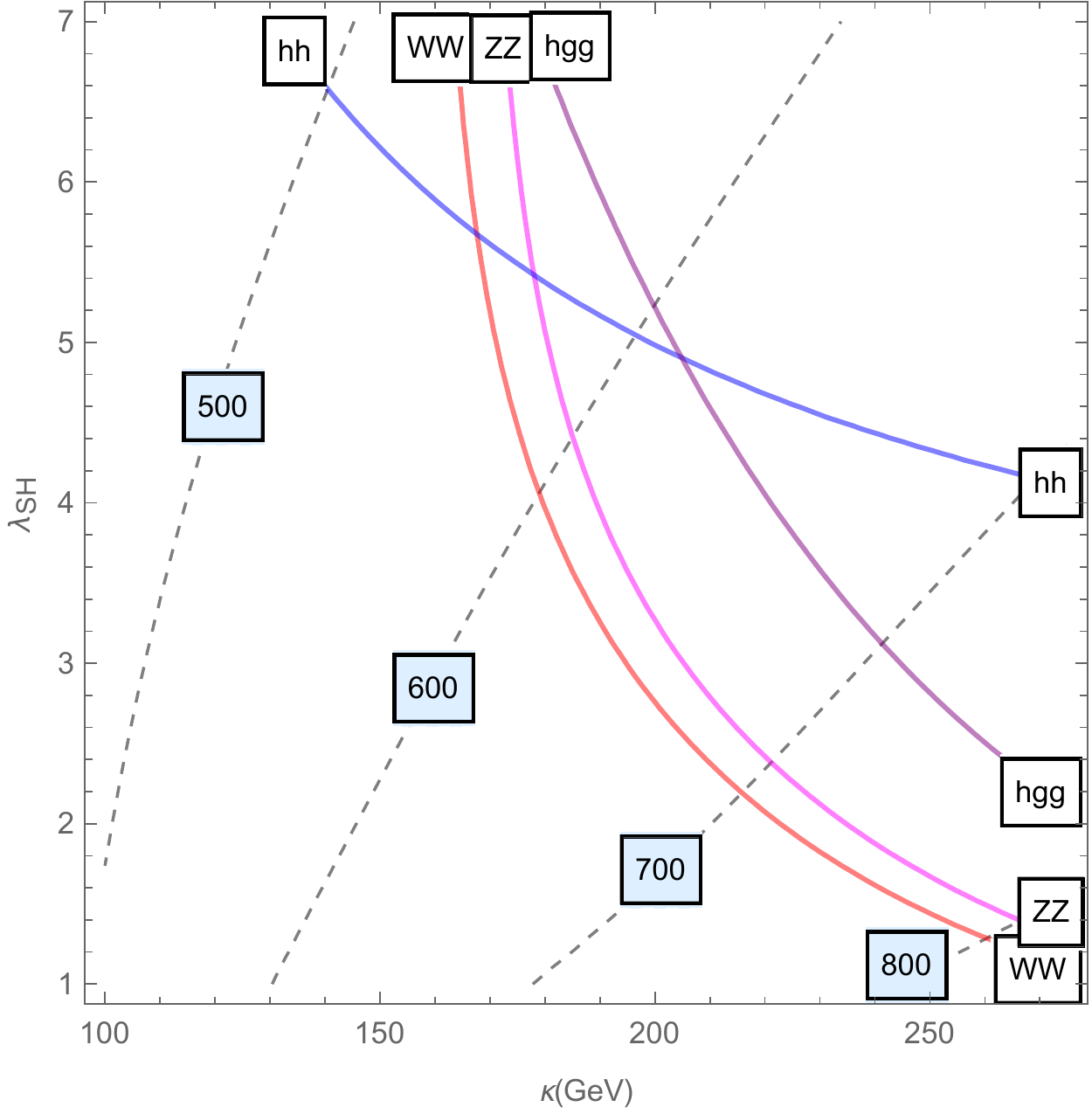}
\includegraphics[width=8cm]{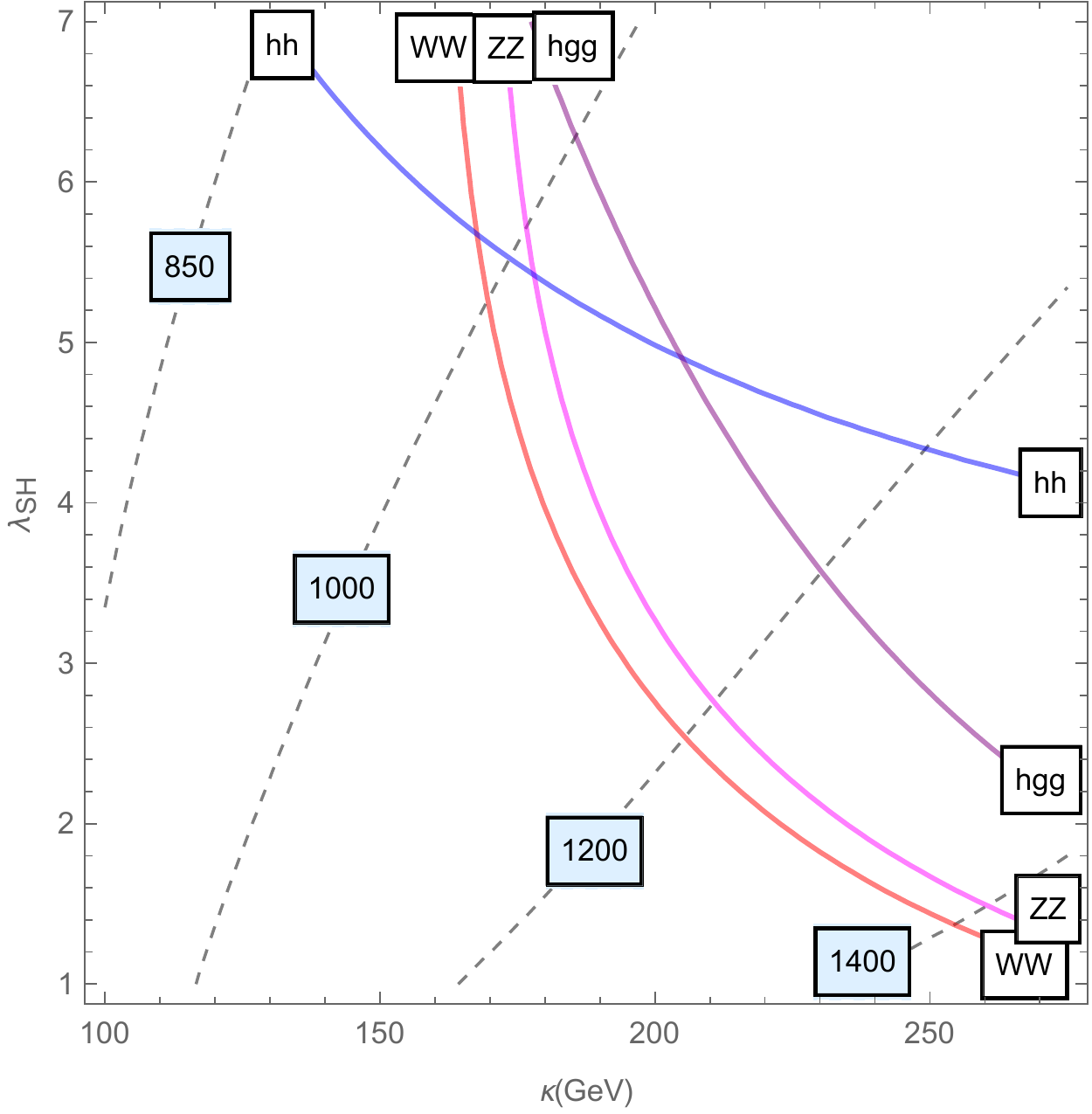}
\vspace{-2mm}
\caption{LHC Run-I constraints on the parameter space of the model (solid lines; the regions above and to the right of the lines are ruled out). Contours of the constant VLQ mass (in GeV) are also shown, with $A=1.0$ (top panel) and $A=0.1$, corresponding to 10\% fine-tuning of the singlet tadpole (bottom panel).}
\label{fig:constraints}
\end{center}
\end{figure}

The constraints imposed by the Run-I data are shown in Fig.~\ref{fig:constraints}. The strongest constraints come from the searches for $W^+W^-$ and di-Higgs resonances at 750 GeV; the $ZZ$ channel and the 125 GeV Higgs decays are currently less constraining. Also shown are the contours of VLQ mass, for $A=1$ (top panel) and $A=0.1$ (bottom panel). For comparison, direct searches for VLQs of charge 5/3 at the LHC Run-I place a lower bound on this mass of $0.75-1.0$ TeV in the simplest models where the VLQ decays into a $W$ boson and an SM quark.   

{\em Electroweak Phase Transition ---} Dynamics of the electroweak phase transition is determined by the effective finite-temperature scalar potential: 
\beq
V_{\rm eff}(S, H; T) = V_{\rm tree}(S, H) + V_{\rm CW}(S, H) + V_T(S, H; T).
\eeq{Veff}
Here $V_{\rm CW}$ is the Coleman-Weinberg potential which includes quantum corrections at zero temperature, while $V_T$ is the finite-temperature effective potential. We compute both $V_{\rm CW}$ and $V_T$ at the one-loop level, including contributions from the SM top and electroweak gauge bosons, as well as $S$ itself. In the case of electroweak gauge contribution, we expand $V_T$ in gauge coupling constants $g$ and $g^\prime$, and truncate the expansion at order $g^2$. This avoids well-known difficulties that stem from gauge-dependence of the full gauge contribution to $V_T$, while being accurate enough for our purposes. (For a more detailed discussion of these issues, see Ref.~\cite{Katz:2015uja}.) For top and scalar-loop contributions to $V_T$, we include the full thermal functions, $J_{B/F}(x)$, without using high-temperature approximations. We also include the ``daisy resummation" terms in $V_T$, which capture the leading corrections beyond the one-loop order~\cite{Fendley:1987ef,Carrington:1991hz}. To study the dynamics of the phase transition
numerically, we adopt the code used previously in Refs.~\cite{Katz:2014bha,Katz:2015uja}. The algorithm computes the effective potential as a function of temperature $T$, starting from low initial $T$ and increasing it in small increments. It searches for the critical temperature $T_c$, defined as the temperature at which $V_{\rm eff}(S, H; T_c)$ has two degenerate local minima, one with $\langle H\rangle=0$ and one with $\langle H\rangle\not=0$. The procedure is repeated for a large set of points in the input parameter space, $(m_Q, \lambda_{SH})$. If the search for $T_c$ yields no result, we conclude that no first-order transition occurs at that point. If the algorithm does find a critical point, we compute 
\beq
\xi = \frac{v(T_c)}{T_c},
\eeq{xi}    
where $v(T_c)$ is the value of $\langle H\rangle$ in the local minimum where it is non-zero. A necessary criterion for electroweak baryogenesis is $\xi\gsim 0.6-0.9$; for smaller values of $\xi$, sphaleron transitions inside the electroweak-broken phase wash out the baryon asymmetry. We will adopt this criterion and refer to points where it is satisfied as having a ``strongly first-order" phase transition. Note that non-zero values of $\langle S\rangle$ at one or both degenerate minima would generally be present in our model. The numerical algorithm automatically accounts for this by searching for local minima in the two-dimensional field space. However, $\langle S\rangle$ does not affect the sphaleron rate and therefore does not enter our criterion for a strongly first-order transition.

\begin{figure}[t!]
\begin{center}
\includegraphics[width=8cm]{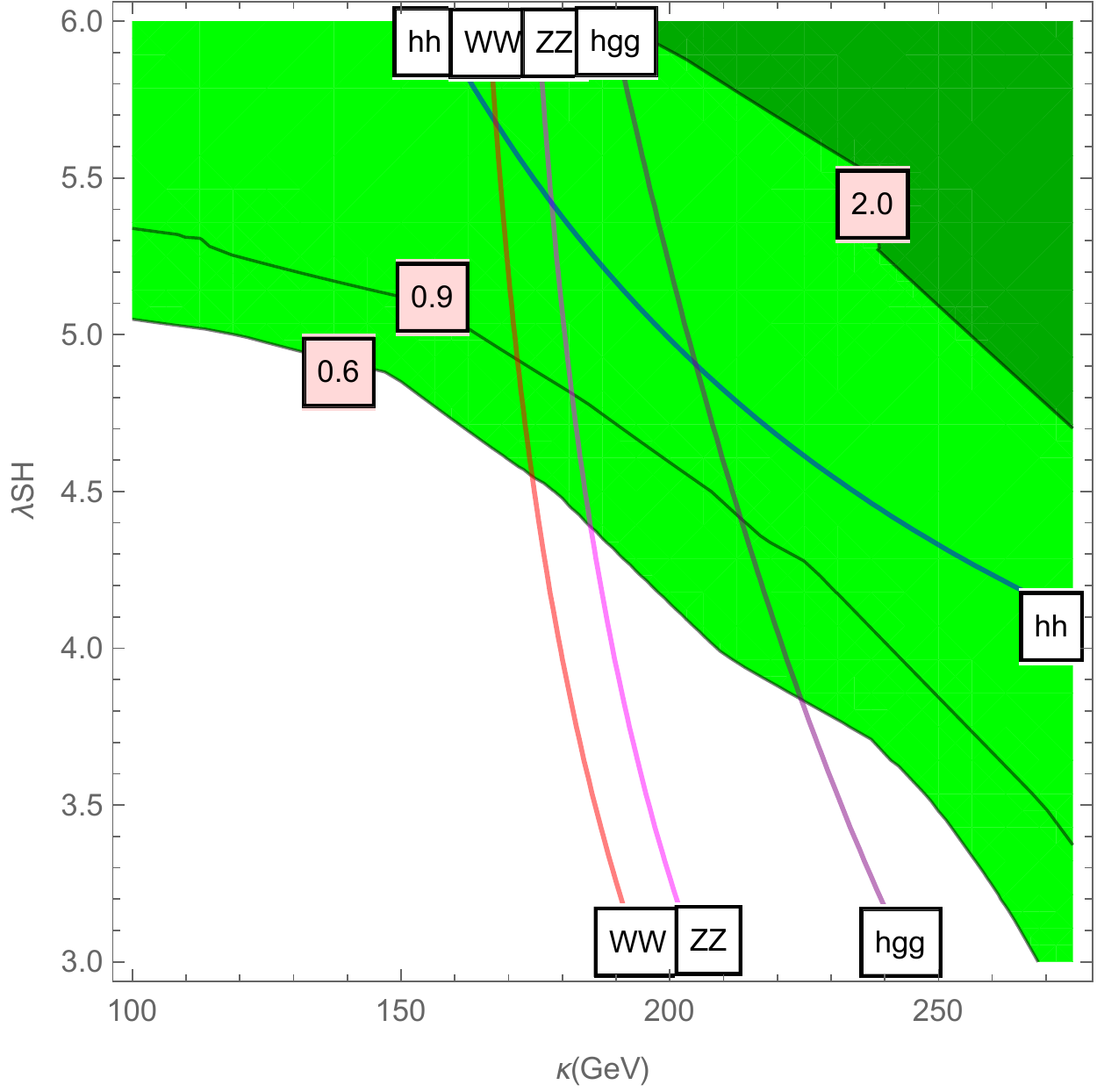}
\vspace{-2mm}
\caption{Regions of strongly first-order electroweak phase transition (green/shaded) along with contours of constant $\xi$. Also shown are the current LHC constraints on the model (solid lines; the regions above and to the right of the lines are ruled out).}
\label{fig:moneyplot}
\end{center}
\end{figure}

The regions in the parameter space with a strongly first-order EWPT are displayed in Fig.~\ref{fig:moneyplot}, together with the constraints from $X$ and $\phi$ decays. The key observation is that there is a region where all these requirements are compatible. In the limit of small $\kappa$, first-order EWPT is achieved for $\lambda_{SH}\gsim 4.5$, which is in good agreement with the analysis of a $Z_2$-symmetric model in Ref.~\cite{Curtin:2014jma}. In the presence of $Z_2$-violating terms in the potential, a first-order EWPT can occur for smaller values of $\lambda_{SH}$. However, given the LHC constraints on the doublet-singlet mixing, the minimum $\lambda_{SH}$ can only be reduced by about 10\%. 

A large value of $\lambda_{SH}$ required for a first-order EWPT raises questions concerning the validity of perturbative analysis. A naive estimate, $\lambda_{SH}\lsim 4\pi$, leaves open a broad range of $\lambda_{SH}$ with a first-order EWPT. This estimate can be refined by considering the singlet contributions to the one-loop Higgs potential, in particular the Higgs quartic coupling renormalization~\cite{Curtin:2014jma}. The minimal values of $\lambda_{SH}$ that produce a first-order EWPT in our model correspond to $\Delta\lambda\sim 0.3-0.4$, which appears to be safely perturbative. Thus, we conclude that the use of perturbation theory in our analysis is justified, although more work would be needed to establish the upper bound on $\lambda_{SH}$ beyond which a non-perturbative approach would be required.

Comparing Figs.~\ref{fig:constraints} and~\ref{fig:moneyplot} indicates that without fine-tuning in the singlet tadpole ($A\sim 1$), the VLQ mass in the region where a first-order EWPT occurs is in the $500-600$ GeV range. This is ruled out in the simplest models, although the bounds may be relaxed in more complex models with non-SM decay channels for VLQ. (For an example of such phenomenology in the case of charge-2/3 top partners, see~\cite{Anandakrishnan:2015yfa}.) If 10\% fine-tuning in the singlet tadpole is assumed, the VLQ mass in the region of interest is about 1 TeV, allowed by the current searches even with conventional VLQ decays. Note that the VLQ mass decreases with decreasing $\kappa$, so that direct VLQ searches imply more fine-tuning at lower $\kappa$. For this reason, the region $\kappa\lsim 100$ GeV (not shown in the plots) is disfavored. The values of the VLQ Yukawa coupling $y_Q$ (see Eq.~\leqn{L_VLQ}) in the region of interest are $\sim 1.0-2.0$ for $A=1$, and $\sim 2.0-4.0$ for $A=0.1$. While higher degree of fine-tuning in the singlet tadpole would further relax the bounds from direct VLQ searches, perturbativity of $y_Q$ serves as a limiting factor, especially since renormalization group evolution leads to further increase in $y_Q$ with energy~\cite{Gu:2015lxj}. 

\begin{figure}[t!]
\begin{center}
\includegraphics[width=8cm]{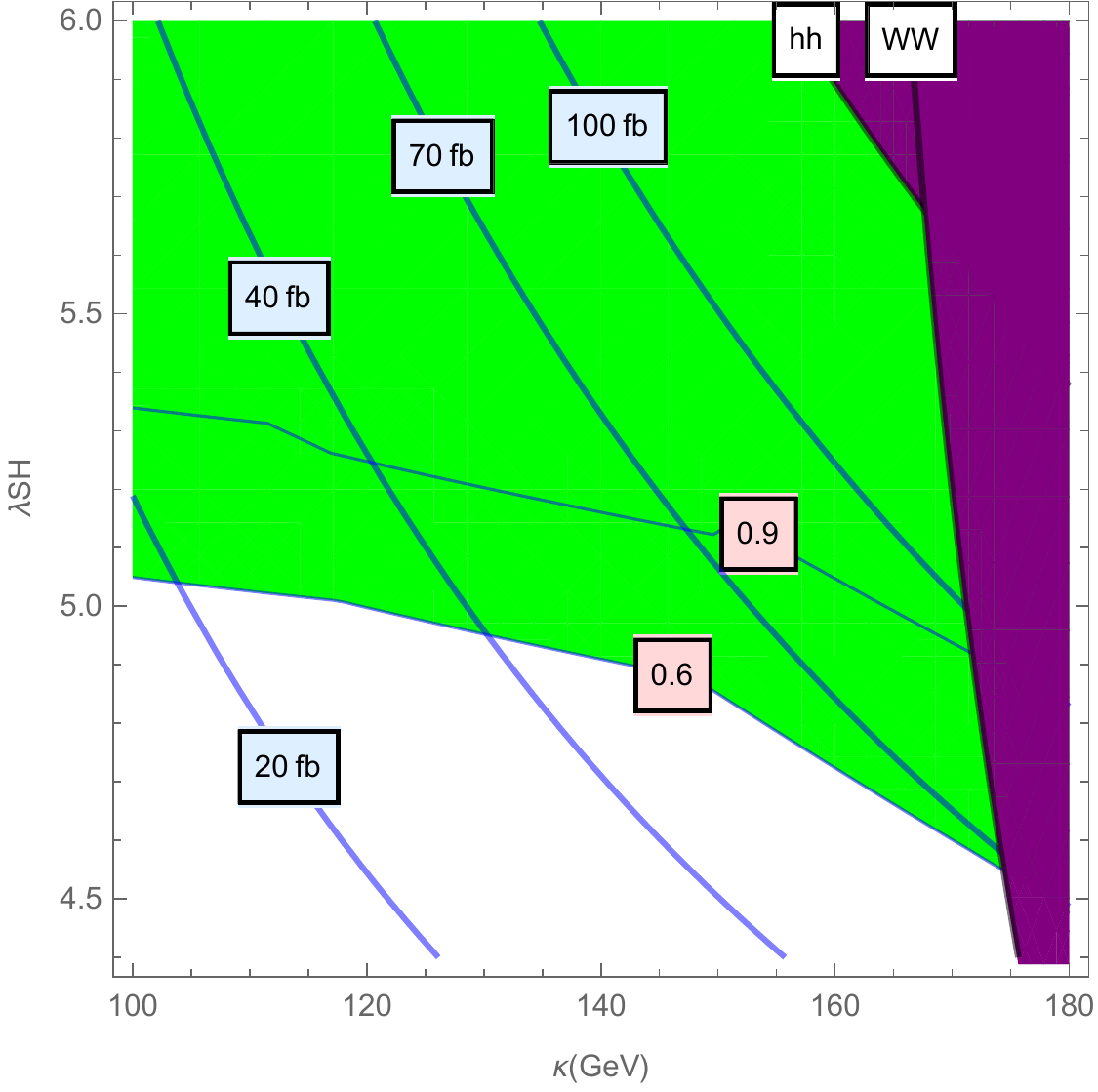}
\vspace{-2mm}
\caption{Cross section of 750 GeV resonant di-Higgs signal in $pp$ collisions at $\sqrt{s}=13$ TeV (solid lines). Also shown are regions of strongly first-order electroweak phase transition (green/light-gray) and the regions ruled out by LHC constraints (purple/dark-grey).}
\label{fig:hh}
\end{center}
\end{figure}

\begin{figure}[t!]
\begin{center}
\includegraphics[width=8cm]{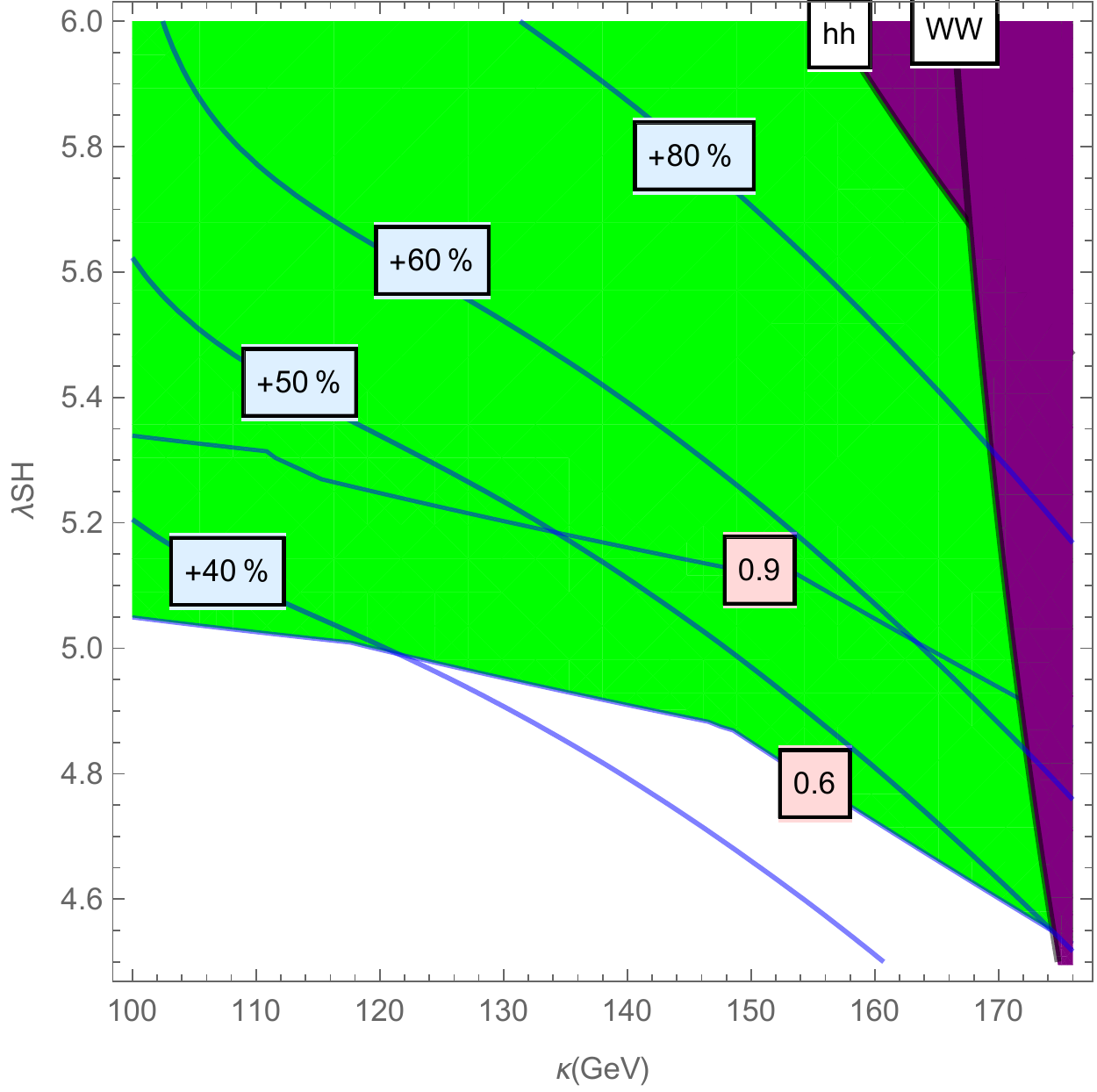}
\vspace{-2mm}
\caption{Fractional deviation of the 125 GeV Higgs cubic self-coupling (solid lines). Also shown are regions of strongly first-order electroweak phase transition (green/light-gray) and the regions ruled out by LHC constraints (purple/dark-grey). }
\label{fig:cubic}
\end{center}
\end{figure}

{\em Predictions for Future Experiments ---} The strongly first-order EWPT scenario in our model produces specific predictions that can be tested in future colliders. Not surprisingly, the observables showing the strongest correlation with the EWPT dynamics are those directly probing the scalar potential. In particular, the cross section for resonant di-Higgs production at 750 GeV at the 13 TeV LHC is $\gsim 20$ fb throughout the parameter space with a first-order EWPT; see Fig.~\ref{fig:hh}. This prediction will be tested very soon. It is, however, somewhat model-dependent, since the rate would be reduced if non-SM decay channels of $S$ are present. A more robust prediction concerns the deviations in the cubic self-coupling of the 125 GeV Higgs boson, which are predicted to be at the level of about 40\% or more, see Fig.~\ref{fig:cubic}. This is yet another example of a generic correlation found in Ref.~\cite{Noble:2007kk}. Deviations of this size can be probed by studying double-Higgs production at the International Linear Collider~\cite{Baer:2013cma,Fujii:2015jha} or a 100 TeV proton collider~\cite{Arkani-Hamed:2015vfh}. Many other predictions of the model -- the presence of VLQs with masses around a TeV, $WW$ and $ZZ$ resonances at 750 GeV, and deviations in the 125 GeV Higgs couplings to photons and gluons, typically at a few-\% level -- will be tested at the LHC Run-II. While these predictions are primarily sensitive to the size of the mixing angle $\theta_{SH}$ and do not show a strong correlation with the first-order EWPT, these searches will of course play a crucial role in testing the underlying framework.

{\em Discussion ---} A 750 GeV gauge-singlet scalar particle, coupled to gluons and photons through loops of vector-like quarks, provides a simple interpretation of the LHC diphoton excess. In this paper, we showed that such a scalar can also lead to a strongly first-order electroweak phase transition, opening the possibility for the electroweak baryogenesis mechanism to be realized. The scenario requires rather large values of the singlet coupling to the Higgs, $\lambda_{SH}\gsim 4.5$. Such couplings naturally arise in composite Higgs models with compositeness scale around a TeV, so that a first-order electroweak phase transition may in fact be a rather generic feature of such models. (Of course, explaining the relatively low mass and small quartic coupling of the Higgs doublet is more challenging in this context.) 

Beyond a first-order EWPT, electroweak baryogenesis scenario requires that additional sources of CP violation, beyond that present in the SM, be introduced around the weak scale. For example, this may be achieved if additional pseudo-scalar states are present in the spectrum, as would typically be the case in composite Higgs models. A detailed analysis of possibilities for CP-violation in this context would be well motivated.  

The mechanism for strongly first-order EWPT considered in this paper will be tested experimentally in the near future. Among the various predictions made by the underlying model, we identified two that appear to be tightly correlated with the EWPT dynamics: the rate of resonant di-Higgs signal at 750 GeV, and the deviation in the Higgs cubic coupling. This highlights the power of near-future experiments probing the Higgs sector to shed light on the nature of the electroweak phase transition, and by extension on the viability of electroweak baryogenesis mechanism. 

{\em Acknowledgments ---} We are grateful to Chian-Shu Chen, Chuan-Hung Chen, Csaba Csaki, Yuval Grossman, Kiel Howe, Eric Kuflik, Salvator Lombardo, Siddharth Mishra-Sharma, Wee Hao Ng and Nicolas Rey-Le Lorier  for useful discussions. We are especially grateful to Andrey Katz and Gowri Kurup for contributions to the numerical code used in the electroweak phase transition analysis. This work is supported by the U.S. National Science Foundation through grant PHY-1316222. YT is also supported by Taiwan Study Abroad Scholarship.

\begin{center}
{\bf Appendix: An Explicit Model with No Tree-Level $Z_2$ Violation in the Scalar Potential}
\end{center}

In this paper, we assumed that the tree-level scalar potential respects the $Z_2$ symmetry, even though this symmetry is broken by the VLQ mass. We envision that the symmetry breaking is spontaneous. The simplest model would be to introduce an additional scalar field $S_1$, odd under the $Z_2$, coupled via
\beq
{\cal L} \subset \ y_Q^\prime S_1 Q_L^\dagger Q_R + y_Q S Q_L^\dagger Q_R +~{\rm h.c.}
\eeq{L_VLQ_ext}   
If $S_1$ gets a vev, this interaction reproduces Eq.~\leqn{L_VLQ}, with $m_Q=y_Q^\prime \langle S_1\rangle$. However, in this simple model, couplings such as $S_1 S$, $S_1 S^3$, and $S_1 S |H|^2$, are all allowed, and should generically be present in the tree level with unsuppressed coefficients. After $S_1$ gets a vev, these terms would generate unacceptable tree-level breaking of $Z_2$. As mentioned in the text, avoiding such terms requires extending this simple model. In this Appendix, we present an explicit example of how this can be achieved. 

First, consider a model with an extended discrete symmetry, $Z_2\times Z_2$. The model contains three scalar fields, with the following discrete symmetry assignments: $S (-, +); S_1 (+, -);  S_2 (-, +)$. In addition there are four Weyl fermions: $Q_L (+, +); Q_R (-, +); U_L (+, -); U_R (+, +)$. The Lagrangian consistent with these symmetries has the form $S Q_L Q_R + m Q_L U_R + S_1 U_L U_R + S_2 U_L Q_R$. When $S_1$ and $S_2$ acquire vevs, all fermions get masses, and $S$ couples to both mass eigenstates with ${\cal O}(1)$ couplings, reproducing the structure assumed in the paper. In the scalar potential, no terms that would generate $S^3$ or $S|H|^2$ are allowed by the extended discrete symmetry. However, in this example, there is still an $S$ tadpole, from $S S_1 S_2$. 

\begin{table}
\centering
\begin{tabular}{l|ll}
Field & $Z_2$ Charges \\ \hline
 $S$ & $-++++$  \\
$S_1$  & $+-+++$ \\
 $S_2$ & $++-++$ \\ 
 $S_3$ & $+++-+$ \\ 
 $S_4$ & $++++-$ \\ 
 $S_5$ & $-----$ \\ 
 $Q_L$ & $+++++$ \\ 
 $Q_R$ & $-++++$ \\ 
 $U_L$ & $-++-+$ \\ 
 $U_R$ & $+-+++$ \\ 
 $B_L$ & $+--++$ \\ 
 $B_R$ & $-++--$ \\ 
 \hline
\end{tabular}
\caption{Field content and charge assignments of a model where no tree-level $Z_2$-breaking terms in the scalar potential are allowed.} 
\label{tab:monster}
\end{table}

Forbidding the $S$ tadpole requires further extending the field content and the symmetry structure of the model. As a proof-of-principle demonstration that there are no fundamental obstacles to doing so, consider a model with a $Z_2^5$ discrete symmetry, whose field content is shown in Table~\ref{tab:monster}. The Lagrangian has the form $S Q_L Q_R + S_1 Q_L U_R + S_2 B_L U_R + S_3 U_L Q_R + S_4 U_L B_R + S_5 B_L B_R$. Again, after all five $S_i$ fields acquire vevs, all fermions get masses, and $S$ couples to all three mass eigenstates with ${\cal O}(1)$ couplings. At the same time, the lowest-order term in the scalar potential which contains an odd power of $S$ and is consistent with the discrete symmetry is a dimension-6 operator $S S_1 S_2 S_3 S_4 S_5$. The coefficient of the $S$ tadpole generated by this term after $S_i$ acquire vevs  is suppressed by the ratio of the vevs and the scale at which this interaction is generated, which can be arbitrarily high. Therefore, the tree-level tadpole and other $Z_2$-breaking terms in this model are naturally small.

\end{document}